\documentclass{article}
\usepackage{hiph-art}
\usepackage{graphicx}
\usepackage{amssymb}
\usepackage{amsmath}
\volnumber{19} \issuenumber{1} \edyear{2005}                             
\frompage{000} \topage{000}                                              
\recrevdate{1 January 2005}                                              
\def\rightmark{Electromagnetic Probes}
\def\leftmark{Thomas, Gallmeister, Zschocke, K\"ampfer}
\title{Electromagnetic Probes of Strongly Interacting Matter:
Probes of Chiral Symmetry Restoration?} 
\authors{ 
{R. Thomas$^1$, K. Gallmeister$^2$, S. Zschocke$^1$, B. K\"ampfer$^1$%
\index{K\"ampfer, B.} 
\index{Thomas, R.} 
\index{Zschocke, S.}
\index{Gallmeister, K.}
}\\[2.812mm]
{\normalsize
\hspace*{-8pt}$^1$ Institut f\"ur Kern- und Hadronenphysik, Forschungszentrum Rossendorf,\\ 
PF 510119, D-01314 Dresden, Germany\\[0.2ex]
\hspace*{-8pt}$^2$ Institut f\"ur Theoretische Physik, Universit\"at Giessen,
D-35392 Giessen,\\ Germany
}}
 
\abstract{The QCD sum rule approach to in-medium modifications of
the $\omega$ meson in nuclear matter is reviewed with emphasis of its relation
to 4-quark condensates and chiral symmetry restoration. 
Possible implications of the CB-TAPS experiment for the reaction
$\gamma A \to A' \omega (\to \pi^0 \gamma)$ are sketched and the particularly
important role of di-electron probes, accessible with HADES, is highlighted.
A brief update of a parametrization of the previous dilepton and photon
probes from CERES and WA98 of heavy-ion collisions at CERN-SPS energies is presented.}
  
 
\makeindex
\begin{document}
 
\maketitle

\section{Introduction}\label{intro}

The chiral condensate $\langle \bar q q \rangle$ 
is an order parameter for the spontaneous breaking
of chiral symmetry in the theory of strong interaction 
(cf.~\cite{Rapp_Wambach} for introducing this topic). 
Its role is demonstrated, e.g., by the Gell--Mann-Oaks-Renner relation
$m_\pi^2 \propto - \langle \bar q q \rangle$ 
for the pion mass\footnote{The analysis of precision data has shown 
that more than 94\% of the pion mass stem from the chiral condensate \cite{Colangelo}.} 
or by Ioffe's formula $m_N \propto - \langle \bar q q \rangle$ 
for the nucleon mass.
There is growing evidence that the quark-gluon condensate
$\langle \bar q \sigma_{\mu \nu} G^{\mu \nu} q \rangle \propto \langle \bar q q \rangle$
\cite{qg_cond} 
(with $G_{\mu \nu}$ as chromodynamic field strength tensor)
is another order parameter. The QCD trace anomaly related to
scale invariance breaking gives rise to the gluon condensate 
$\langle (\alpha_s / \pi) G_{\mu \nu} G^{\mu \nu} \rangle$. There are many other condensates
characterizing the complicated structure of the QCD vacuum. In a medium,
described by temperature $T$ and baryon density $n$, these condensates change,
i.e., the ground state is rearranged. Since hadrons are considered as excitations
above the vacuum, a vacuum change should manifest itself as a change of the
hadronic excitation spectrum. This idea triggered widespread activities
to search for such in-medium modifications of hadrons and to relate them
to corresponding order parameters. 

Among the promising tools are electromagnetic probes. Due to the small
cross sections, real or virtual photons can leave nuclear systems nearly
undisturbed and carry information on their creation conditions. This is
evidenced in the realm of heavy-ion collisions by
photon ($\gamma$) and dielectron ($e^+ e^-$)
emission rates
\begin{eqnarray}
\omega \frac{d R_\gamma}{d^3 k} & = & - {\rm Im}\Pi_\mu^\mu \, n_B(k_0/T)
\frac{2}{(2 \pi)^3}, \label{photon_rate}\\
\frac{d R_{e^+ e^-}}{d^4 Q} & = & - {\rm Im}\Pi_\mu^\mu \, n_B(Q_0/T) 
\frac{\alpha}{12 \pi^4 M^2} ,
\label{dilepton_rate}
\end{eqnarray} 
where $\Pi_\mu^\mu$ is the trace of the retarded current-current correlator representing
at the same time the photon self-energy in the medium, and $n_B$ is the Bose distribution
function; $k$ is the photon momentum and $Q$ denotes the 4-vector of the
$e^+ e^-$ pair where the electron mass is neglected and $M$ is the invariant $e^+ e^-$ mass.
With virtue of (\ref{photon_rate}, \ref{dilepton_rate}), real and virtual photons
are ideal probes of the medium properties. Moreover, due to vector dominance
\cite{VDM} photons couple predominantly to vector mesons. Thus, the in-medium
behavior of light vector mesons $\rho$, $\omega, \, \cdots$ becomes directly accessible. 

Concentrating on the iso-scalar part
the causal current-current correlator, here specified for the $\omega$ meson,
\begin{equation}
\Pi^\omega(q,n) = 
\frac i3 \int d^4 x \, e^{iqx} \langle \Omega | {\cal T} j^\omega_\mu (x) {j^\omega}^\mu (0) 
| \Omega \rangle
\end{equation}
with the current 
$j^\omega_\mu = \left( \bar u  \gamma_\mu u +^* \bar d  \gamma_\mu d \right)/2$,\footnote{
Here and afterwards an asterisk indicates a sign change for the $\rho$ meson.}
an operator product expansion and a Borel transformation of the
twice-subtracted dispersion relation results in 
\begin{equation}  
\dfrac{1}{\pi} \int_0^\infty ds \dfrac{{\rm Im} \Pi^\omega(s,n)}{s} e^{-s/{\cal M}^2} 
= - c_0 {\cal M}^2 - \sum_{j=1}^\infty \dfrac{c_j}{(j-1)! {\cal M}^{2(j-1)}}
+ \Pi^\omega (0,n),
\label{sum_rule_0}
\end{equation}
where $\Pi^\omega (0,n) = 9 n/(4 m_N)$\footnote{
$n$ denotes the baryon density and $m_N$ is the nucleon mass.}
is a subtraction constant with the meaning of Landau damping
or $\omega$ forward scattering amplitude, and the coefficients $c_j$ contain condensates
and Wilson coefficients; ${\cal M}$ is the Borel mass. 

Through the chain of equations 
(\ref{photon_rate} - \ref{sum_rule_0}) electromagnetic signals off hadron
systems and QCD condensates are interrelated.
Chiral symmetry restoration, in this context, means a dropping chiral condensate
which modifies the electromagnetic emission pattern \cite{BR}. As mentioned above
there exist many condensates which are modified in the nuclear medium. Therefore,
the resulting picture becomes more complex. As we are going to point out, in truncated
sum rules different hadrons are sensitive to different condensates.

Our paper is organized as follows. In section 2 we discuss in some detail the
QCD sum rule for the $\omega$ meson and try to draw first consequences from
the recent CB-TAPS experiment. We contrast in section 3 the crucial dependence of the 
$\omega$ properties on 4-quark condensates with the QCD sum rules for the nucleon
which is sensitive to the chiral condensate. In section 4 we present a brief
update of the description of data on electromagnetic probes at CERN-SPS energies.
The summary can be found in section 5.   

\section{QCD Sum Rule for $\omega$ Meson}\label{omega meson sum rules}  

The truncated QCD sum rule (\ref{sum_rule_0}) for the $\omega$ meson 
can be arranged as \cite{SZ2004}
\begin{equation}
m_\omega^2 (n,{\cal M}^2,s_\omega) = \dfrac{c_0 {\cal M}^2 
\left[ 1 - \left ( 1 + \dfrac{s_\omega}{{\cal M}^2} \right) e^{-s_\omega / {\cal M}^2} \right] - 
\dfrac{c_2}{{\cal M}^2} - \dfrac{c_3}{{\cal M}^4} - \dfrac{c_4}{2 {\cal M}^6}}
{c_0 \left ( 1 - e^{-s_\omega / {\cal M}^2} \right) + \dfrac{c_1}{{\cal M}^2} 
+ \dfrac{c_2}{{\cal M}^4} + 
\dfrac{c_3}{2 {\cal M}^6} + 
\dfrac{c_4}{6 {\cal M}^8} - 
\dfrac{\Pi^\omega (0,n)}{{\cal M}^2}},
\label{sum_rule}
\end{equation}
where the quantity $m_\omega^2 (n,{\cal M}^2,s_\omega)$ 
is the normalized moment of the imaginary part
of the current-current operator
\begin{equation}
m_\omega^2 (n,{\cal M}^2,s_\omega) = \dfrac{\int_0^{s_\omega} ds \; {\rm Im} \Pi^\omega (s,n) \; 
e^{-s/{\cal M}^2}}{\int_0^{s_\omega} ds \; {\rm Im} \Pi^\omega (s,n) \; s^{-1} e^{-s/{\cal M}^2}}.
\label{mass_parameter}
\end{equation}
Sum rules, even with resonance + continuum ansatz as exploited in 
(\ref{sum_rule}, \ref{mass_parameter}),
constrain some integral strength but not individual parameters without
further specification of ${\rm Im} \Pi$.   

The first coefficients $c_j$ in (\ref{sum_rule_0}) refer to the decomposition       
$\Pi^\omega (Q^2 = - q_\mu q^\mu) = \Pi^\omega_{\rm scalar} + \Pi^\omega_{d = 4,\tau = 2} 
+ \Pi^\omega_{d = 6,\tau = 2} + \Pi^\omega_{d = 6,\tau = 4} + \ldots$ 
in the 2-flavor sector with
\begin{eqnarray}
\Pi^\omega_{\rm scalar}
&=& - \dfrac{1}{8\pi^2} \left( 1 + \dfrac{\alpha_s}{\pi} \right) Q^2 \ln \dfrac{Q^2}{\mu^2}
- \dfrac{3}{8\pi^2} (m_u^2 + m_d^2) \\
&& + \dfrac{1}{2} \left( 1 + \dfrac{\alpha_s}{3\pi} \right) \dfrac{1}{Q^2} 
\left( m_u \langle \bar u u \rangle + m_d \langle \bar d d \rangle \right)
+ \dfrac{1}{24} \dfrac{1}{Q^2} \langle \dfrac{\alpha_s}{\pi} G^2 \rangle \nonumber \\
&& - \dfrac12 \pi \alpha_s \dfrac{1}{Q^4} 
\left( \langle \bar u \gamma_\mu \gamma_5 \lambda^a u \bar u \gamma^\mu \gamma_5 \lambda^a u \rangle
+ \langle \bar d \gamma_\mu \gamma_5 \lambda^a d \bar d \gamma^\mu \gamma_5 \lambda^a d \rangle \right)
\nonumber \\
&& -^* \pi \alpha_s \dfrac{1}{Q^4} 
\langle \bar u \gamma_\mu \gamma_5 \lambda^a u \bar d \gamma^\mu \gamma_5 \lambda^a d \rangle 
- \dfrac29 \pi\alpha_s \dfrac{1}{Q^4} 
\langle \bar u \gamma_\mu  \lambda^a u \bar d \gamma^\mu \lambda^a d \rangle 
\nonumber\\
&& - \dfrac19 \pi \alpha_s \dfrac{1}{Q^4} \left( 
\langle \bar u \gamma_\mu \lambda^a u \bar{u} \gamma^\mu \lambda^a u \rangle 
+ \langle \bar d \gamma_\mu \lambda^a d \bar{d} \gamma^\mu \lambda^a d \rangle \right) \nonumber \\
&& + g_s \dfrac{1}{12} \dfrac{1}{Q^6} \left( m_u^2 \langle  m_u \bar u \sigma_{\mu\nu} G^{\mu\nu} u \rangle 
+ m_d^2 \langle m_d \bar{d} \sigma_{\mu\nu} G^{\mu\nu} d \rangle \right), \nonumber 
\end{eqnarray}
\begin{eqnarray}
\Pi^\omega_{d=4, \tau = 2} 
&=& \dfrac{\alpha_s}{\pi} \dfrac{1}{Q^4} q^\mu q^\nu \langle \hat{S} \hat{T} 
\left( G_\mu^{\;\;\alpha} G_{\alpha\nu} \right) \rangle \\
&& - \left( \dfrac{2}{3} - \dfrac{10\alpha_s}{27\pi} \right) i \dfrac{1}{Q^4} q^\mu q^\nu 
\langle \hat{S} \hat{T} \left( \bar u \gamma_\mu D_\nu u 
+ \bar d \gamma_\mu D_\nu d \right) \rangle, \nonumber \\
\Pi^\omega_{d=6, \tau = 2} 
&=& -\dfrac{82\alpha_s}{27\pi} \dfrac{1}{Q^8} q^\mu q^\nu q^\lambda q^\sigma 
\langle \hat{S} \hat{T} \left( G_\mu^{\;\;\rho} D_\nu D_\lambda G_{\rho\sigma} \right) \rangle \\
&& + \left( \dfrac{8}{3} + \dfrac{134\alpha_s}{45\pi} \right) 
i \dfrac{1}{Q^8} q^\mu q^\nu q^\lambda q^\sigma 
\langle \hat{S} \hat{T} \left( \bar u \gamma_\mu D_\nu D_\lambda D_\sigma u 
+ \bar d \gamma_\mu D_\nu D_\lambda D_\sigma d \right) \rangle, \nonumber\\
\Pi^\omega_{d=6, \tau = 4} 
&=& -^* \dfrac{1}{3} \dfrac{1}{Q^6} q^\mu q^\nu \langle g_s^2 \hat{S} \hat{T} 
\left( \bar u \gamma_\mu \gamma_5 \lambda^a u \bar d \gamma_\nu \gamma_5 \lambda^a d \right)\rangle \\
&& - \dfrac{1}{6} \dfrac{1}{Q^6} q^\mu q^\nu \langle g_s^2 \hat{S} \hat{T} 
\left( \bar u \gamma_\mu \gamma_5 \lambda^a u \bar u \gamma_\nu \gamma_5 \lambda^a u 
+ \bar{d} \gamma_\mu \gamma_5 \lambda^a d \bar{d} \gamma_\nu \gamma_5 \lambda^a d \right) \rangle 
\nonumber \\
&& - \dfrac{1}{24} \dfrac{1}{Q^6} q^\mu q^\nu \langle g_s^2 \hat{S} \hat{T} 
\left( \bar u \gamma_\mu \lambda^a u \left ( \bar u \gamma_\nu \lambda^a u 
+ \bar{d} \gamma_\nu \lambda^a d \right ) \right ) \rangle \nonumber \\
&& - \dfrac{1}{24} \dfrac{1}{Q^6} q^\mu q^\nu \langle g_s^2 \hat{S} \hat{T} 
\left( \bar d \gamma_\mu \lambda^a d \left ( \bar u \gamma_\nu \lambda^a u 
+ \bar{d} \gamma_\nu \lambda^a d \right ) \right ) \rangle \nonumber \\
&& - \dfrac{5}{12} \dfrac{1}{Q^6} q^\mu q^\nu \langle i g_s \hat{S} \hat{T} 
\left( \bar u \left[ D_\mu , \tilde{G}_{\nu\alpha} \right]_+ \gamma^\alpha \gamma_5 u 
+ \bar d \left[ D_\mu , \tilde{G}_{\nu\alpha} \right]_
+ \gamma^\alpha \gamma_5 d \right ) \rangle \nonumber \\
&& - \dfrac{7}{3} \dfrac{1}{Q^6} q^\mu q^\nu \langle g_s^2 \hat{S} \hat{T} 
\left( m_u \bar u D_\mu D_\nu u + m_d \bar d D_\mu D_\nu d \right) \rangle \nonumber
\end{eqnarray}
($\langle \cdots \rangle$ is a short hand notation for
the Gibbs average over the nuclear matter states
$\langle \Omega \vert \cdots \vert \Omega \rangle$,
$\hat S \hat T$ denote operators causing symmetric and traceless structures,
$\alpha_s = g_s^2 / 4\pi$, and $D_\mu$ denotes the covariant derivative;
for further details cf.\ \cite{SZ2004,Oleg}). 
We have spelled out these first terms to evidence that even in low order
much more (and partially poorly known) condensates determine the sum rule.
The genuine chiral condensates 
$\langle \bar u u \rangle$ and $\langle \bar d d \rangle$ enter $c_2$ with the small
factors $m_{u,d}$ and are irrelevant, both in vacuum and in nuclear matter. This is 
exhibited in Fig.~1, where the influence of various terms is discussed.
In most previous approaches the 4-quark condensates have been factorized,
i.e., expressed by square of the chiral condensate. Within such an approximation,
statements that the in-medium change of the chiral condensate drives in-medium
modifications of light vector mesons are correct. However, strictly speaking
it is the change of 4-quark condensates which drive the change of $m_\omega$,
together with other subleading terms such as the gluon condensate.

\begin{figure}[htb]
\vspace*{0.1cm}
\includegraphics[width=6.3cm]{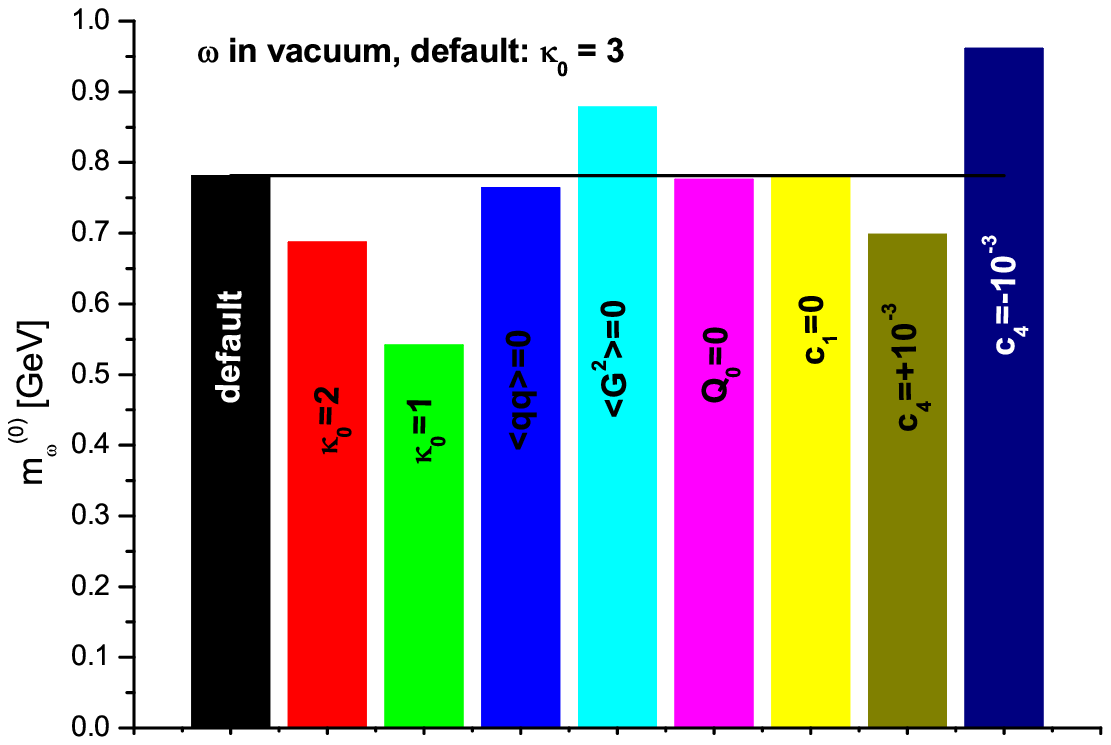}
\hfill
\includegraphics[width=6.3cm]{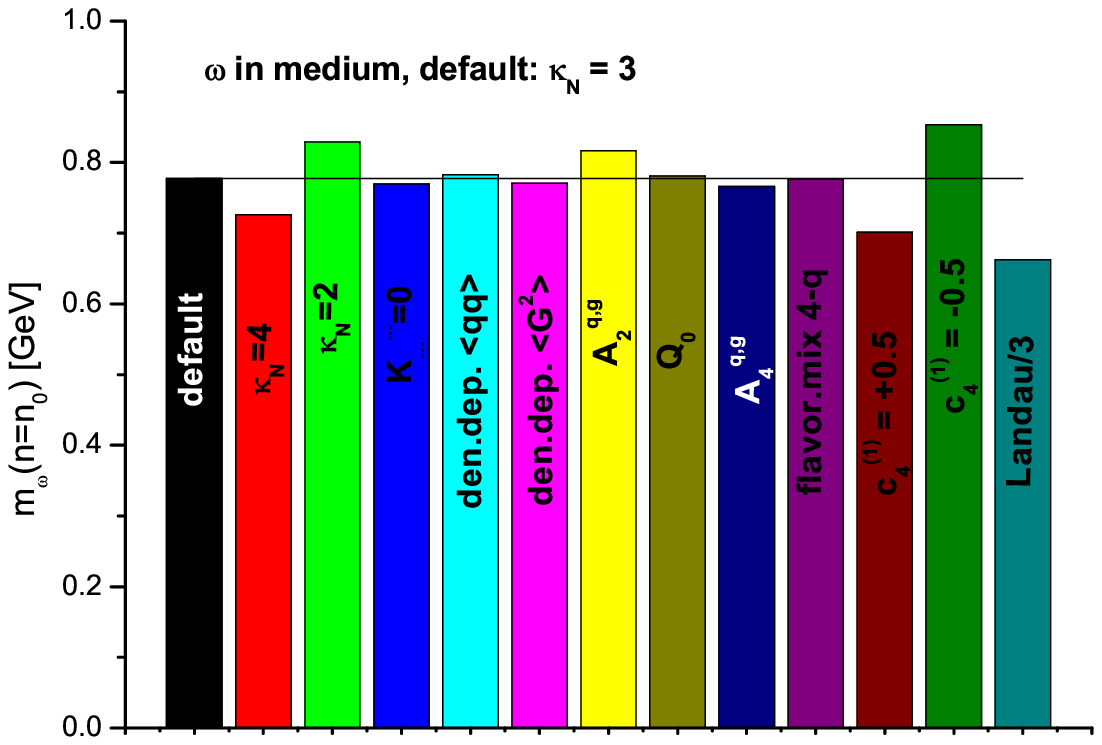}
\vspace*{-0.6cm}
\caption[]{Importance of various quantities for the parameter $m_\omega$
in the QCD sum rule (\ref{sum_rule}). Left (right) panel: for vacuum (nuclear matter).
Default values are $\langle \bar q q \rangle_0 = (-0.245 \; \mathrm{GeV})^3$ and
$\langle (\alpha_s/\pi) G^2 \rangle_0 = (0.33 \; \mathrm{GeV})^4$. 
The 4-quark condensates enter $c_3$ in the combination
$-\frac{112}{81} \pi \alpha_s \kappa_0 \langle \bar q q \rangle_0^2
[1 + \frac{\kappa_N}{\kappa_0} \frac{\sigma_N}{m_q \langle \bar q q \rangle_0} n]$
with $\sigma_N = 45$ MeV and $m_q = 5.5$ MeV. 
Furthermore, $c_3$ contains a term $\propto Q_0^2$ responsible for the $\rho - \omega$
mass splitting in vacuum. $A_{2,4}^{q,g}$ are contributions to $c_3$ which can be traced back
to moments of the parton distribution in the nucleon.
''flavor.mix'' and $K_{\cdots}^{\cdots}$ indicate poorly known terms in $c_3$
stemming from flavor mixing condensates.
The term $c_1$ is determined by $m_{u,d}^2$.
The complete coefficient $c_4$
is presently unknown. In the default calculation this term is discarded. To estimate its
influence we assume {\it ad hoc} 
$c_4(n) = \pm 10^{-3} \mathrm{GeV}^8(1+c_4^{(1)} n)$
with $c_4^{(1)} = \pm 0.5 / n_0$.}
\label{fig1}
\end{figure}

\begin{figure}[htb]
\vspace*{-0.3cm}
\centerline{\includegraphics[width=6.0cm]{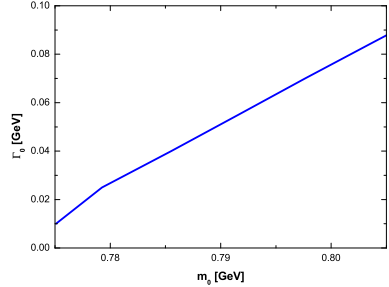}}
\vspace*{-0.1cm}
\caption[]{Emerging correlation $\Gamma_0 (m_0)$ from (\ref{mass_parameter})
when assuming $m_\omega = 0.77$ GeV. The plot is for ${\cal M}=1$ GeV and $s_0=0.135$ GeV.}
\label{fig_gamma_m}
\end{figure}

It should be emphasized that most sum rule approaches calculate the parameter
$m_\omega^2$ via (\ref{sum_rule}) by requiring independence of the Borel mass
${\cal M}$ by choosing an appropriate continuum threshold $s_\omega$ 
and identify  $m_\omega$ with the pole mass of 
the physical particle. 
This is justified in zero-width approximation. In general, however, one needs
a model or data for deducing conclusions on ${\rm Im}\Pi$. To illustrate
this feature we choose as model of ${\rm Im}\Pi^\omega$ a Breit-Wigner distribution,
i.e., ${\rm Im}\Pi = s^{1/2} \Gamma(s) [(s-m_0^2)^2 + s \Gamma(s)^2]^{-1}$
with $\Gamma(s)=\Gamma_0 \sqrt{1-(s_0/s)^2}/\sqrt{1-(s_0/m_0^2)^2}$.
We assume that the sum rule (\ref{sum_rule}) gave $m_\omega = 0.77$ GeV.
Then the expression (\ref{mass_parameter}) determines the correlation
$\Gamma_0 (m_0)$ as exhibited in Fig.~\ref{fig_gamma_m}. Such a correlation
has been noted for the first time in \cite{Leupold_Mosel}.
This example shows that the sum rule approach needs further ingredients
to pin down interesting parameters of ${\rm Im} \Pi$. 
Another possibility to employ the sum rule is to determine $m_\omega$ from
data via (\ref{mass_parameter}) with experimental input ${\rm Im}\Pi^\omega$.
In this respect the electromagnetic probes are superior, as ${\rm Im}\Pi$
is directly determined by rates, see (\ref{photon_rate}, \ref{dilepton_rate}).

The CB-TAPS collaboration \cite{Trnka} has studied the reaction
$\gamma A \to A' \omega (\to \pi^0 \gamma)$. Count rates as a function of
the invariant mass $m_{\pi^0 \gamma}$ have been measured for $LH_2$ and $Nb$
targets. It is remarkable that for the $Nb$ target the strength is noticeably
shifted to smaller values of $m_{\pi^0 \gamma}$ for small values of the
$\omega$ momentum, while for larger $\omega$ momenta the difference for
both targets diminishes. Even without knowledge of
acceptance corrections, phase space factor etc.\ one may use the count rate
as estimator of ${\rm Im}\Pi^\omega$\footnote{
Indeed, the rate $\omega \to \pi^0 \gamma$ is given by
$dR_{\omega \to \pi^0 \gamma} /d^4q \propto \Phi (q) \, {\rm Im}\Pi^\omega (q)$,
where $\Phi$ is a phase space factor. The absolute normalization
cancels in (\ref{mass_parameter}).}
and deduce a small down shift of $m_\omega$ from (\ref{mass_parameter}).
Supposing (i) the validity of the sum rule (\ref{sum_rule}) as it stands, and
(ii) assuming that the measured decays $\omega \to \pi^0 \gamma$
happen to a large extent in nuclear matter, 
and (iii) neglecting poorly known higher terms,
the unavoidable conclusion is that in-medium
4-quark condensates are reduced at least by 30\%. 
(The actual number depends on the used value of $m_\omega$ and details of the chosen
Borel window. We here assumed that the in-medium $\omega$ mass is 0.77 GeV, as in vacuum.)
Such a strong density dependence of
4-quark condensates is needed to counterbalance the strong Landau damping \cite{Hoffmann}
which tends to up-shift the $\omega$ mass parameter in nuclear matter.
One severe problem is the final state interaction
of the $\pi^0$ which may cause a diminished value of $m_{\pi^0 \gamma}$
even when the parent $\omega$ meson has its vacuum mass \cite{Muhlich}.
Nevertheless, the fact that the in-medium $\omega$ mass parameter is not larger
than in vacuum has the mentioned consequence for the 4-quark condensate.
More quantitative results are possible for the reaction
$\pi A \to A' \omega (\to e^+ e^-)$ as envisaged in the HADES physics program
\cite{HADES}:
Due to a similar relation as (\ref{dilepton_rate}), ${\rm Im} \Pi$ is
directly accessible without facing the problem of final state interaction.

\section{QCD Sum Rules for the Nucleon}

\begin{figure}[htb]
\vspace*{-0.1cm}
\includegraphics[width=4.5cm,angle=90]{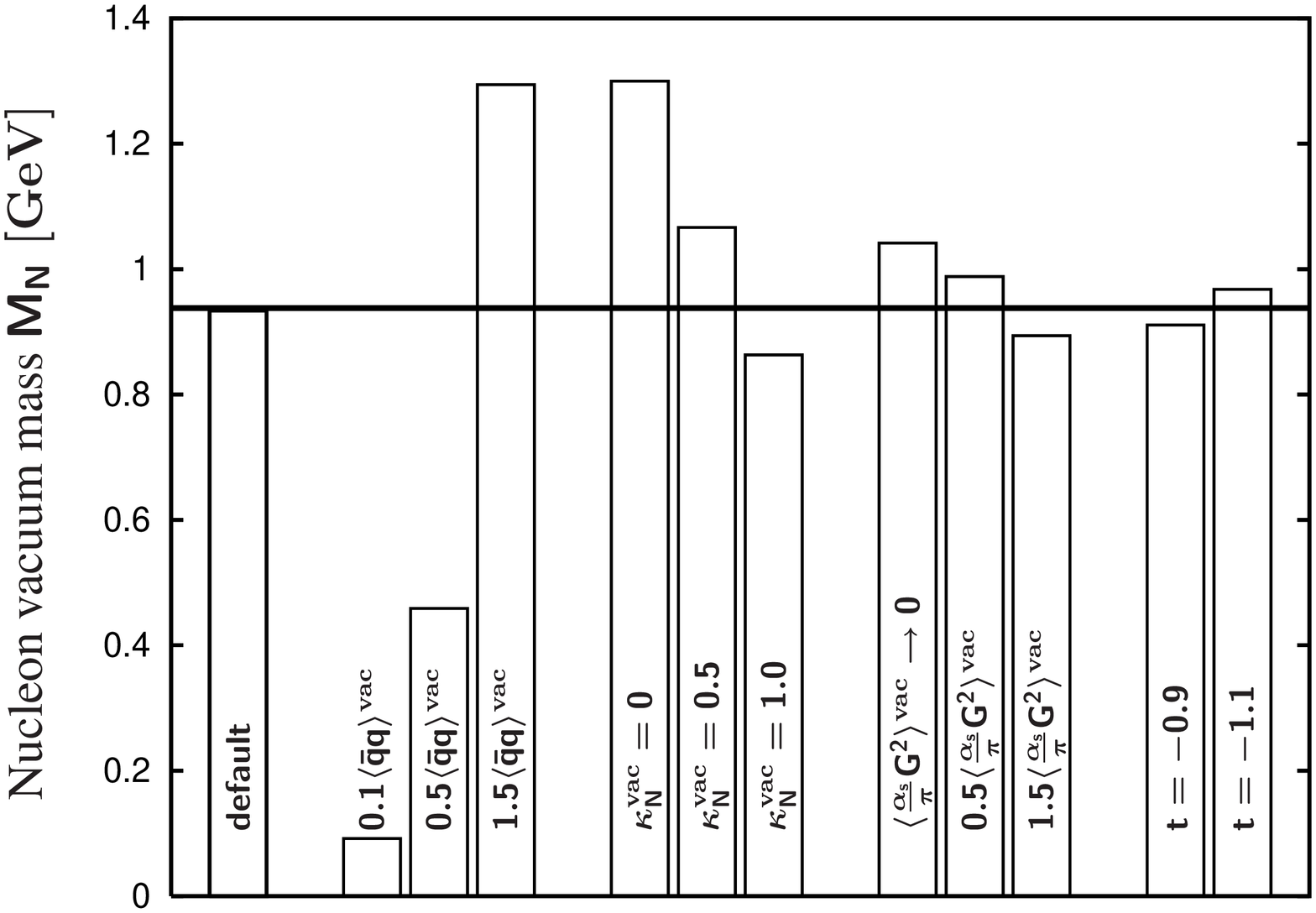}
\hfill
\includegraphics[width=4.5cm,angle=90]{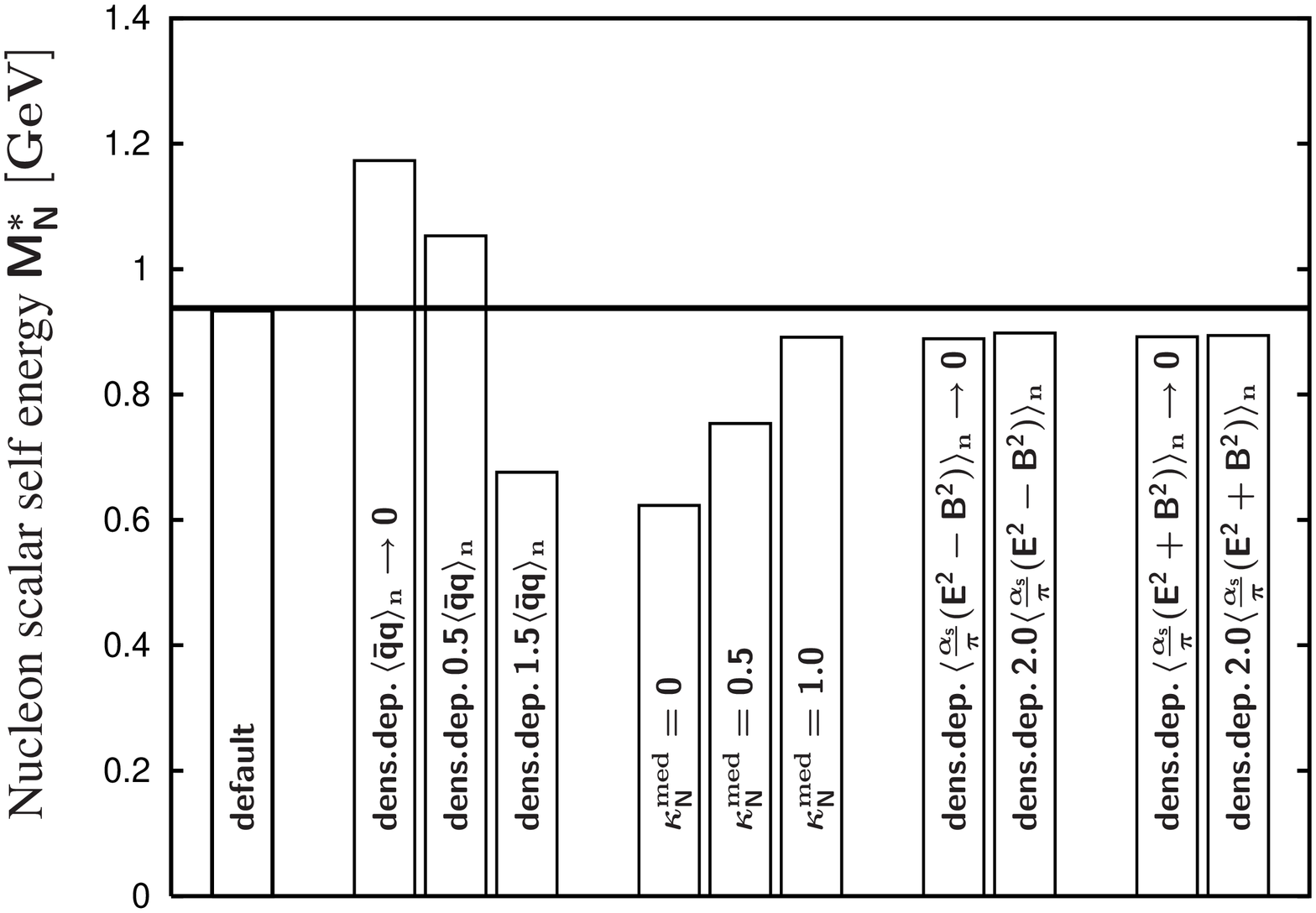}
\vspace*{-0.01cm}
\caption[]{Importance of various contributions to the nucleon mass.
Default values are as for the $\omega$ meson, and the interpolating fields have
a default mixing angle of $t=-1$. The factorized and summed up 4-quark condensates are
fine-tuned with a factor $\kappa_N^{vac} = 0.8$ to the vacuum nucleon mass.
Left panel: Nucleon in vacuum. 
Right panel: Nucleon in medium at saturation density.
The linear density dependence of in-medium 4-quark condensates 
is regulated by the factor $\kappa_N^{{\rm med}}$, similar to the case of $\omega$.
The variations of gluon condensate, whose density dependence is written here in terms
of the color-electromagnetic fields $E$ and $B$, cause minor effects.}
\label{fig2}
\end{figure}

In contrast to the QCD sum rule for the light vector mesons, the chiral condensate
is an important quantity for the nucleon mass. The detailed reminder of the
QCD sum rules for the nucleon will be published elsewhere. We present here only
a few of the striking results to expose the interplay of chiral condensate
and 4-quark condensates. In the left panel of Fig.~\ref{fig2} 
the dependence of the nucleon mass
in vacuum on various quantities entering the sum rule is exhibited. 
One observes a nearly linear dependence on the genuine chiral condensate 
supporting the validity of the Ioffe formula. The impact of 4-quark
condensates is sizeable (their strength is scaled by $\kappa_N^{{\rm vac}}$),
while the gluon condensate is of minor relevance.
The in-medium changes of the chiral condensate cause substantial changes
of the scalar nucleon self-energy, see right panel of Fig.~\ref{fig2}.
Further, the density dependence of the 4-quark condensates
parameterized by a strength factor $\kappa_N^{{\rm med}}$
becomes important in the nuclear medium. 
As noticed in \cite{Furnstahl}, the known
phenomenology of the in-medium nucleon is obtained only by assuming
a weak density dependence of the 4-quark condensates. 
This looks like a contradiction to results of the
previous section. It must be noted, however, that the nucleon sum rule is
entered by many more 4-quark condensates. In Table I we list for the vacuum
the possible 4-quark condensates and their weight in $\omega$
and nucleon sum rules. It is conceivable that the net effect of the 4-quark condensates
leads to some cancellation for the nucleon, while the $\omega$ meson
sum rule is sensitive to the few entering 4-quark condensates. For a calculation
of the 4-quark condensates for the nucleon within a chiral quark model we refer
the interested reader to \cite{Tuebingen}.       

\begin{table}
~\center
\begin{tabular}{lrcc}
\hline
\\
Condensate & Factorized Form & \multicolumn{2}{c}{Contribution for}\\
& & $\omega$ Meson & Proton \\
\hline
\\
$\langle \bar u u \bar{u} u \rangle $ & $11/12 \langle \bar{u} u \rangle ^2$ & 0 & 1/2\\
$\langle \bar u \gamma^\mu u \bar{u} \gamma_\mu u \rangle $ & $-1/3 \langle \bar{u} u \rangle ^2$ & 0 & 1/4\\
$\langle \bar u \sigma^{\mu\nu} u \bar{u} \sigma_{\mu\nu} u \rangle $ & $-1 \langle \bar{u} u \rangle ^2$ & 0 & 0\\
$\langle \bar u \gamma_5 \gamma^\mu u \bar{u} \gamma_5 \gamma_\mu u \rangle $ & $1/3 \langle \bar{u} u \rangle ^2$ & 0 & -1/4\\
$\langle \bar u \gamma_5 u \bar{u} \gamma_5 u \rangle $ & $ -1/12 \langle \bar{u} u \rangle ^2$ & 0 & -1/2\\
\\
$\langle \bar u \lambda^A u \bar{u} \lambda^A u \rangle $ & $-4/9  \langle \bar{u} u \rangle ^2$ & 0 & -3/8\\
$\langle \bar u \gamma^\mu \lambda^A u \bar{u} \gamma_\mu \lambda^A u \rangle $ & $-16/9 \langle \bar{u} u \rangle ^2$ & 2/9 & -3/16\\
$\langle \bar u \sigma^{\mu\nu} \lambda^A u \bar{u} \sigma_{\mu\nu} \lambda^A u \rangle $ & $ -16/3 \langle \bar{u} u \rangle ^2$ & 0 & 0\\
$\langle \bar u \gamma_5 \gamma^\mu \lambda^A u \bar{u} \gamma_5 \gamma_\mu \lambda^A u \rangle $ & $16/9 \langle \bar{u} u \rangle ^2$ & 1 & 3/16\\
$\langle \bar u \gamma_5 \lambda^A u \bar{u} \gamma_5 \lambda^A u \rangle $ & $-4/9  \langle \bar{u} u \rangle ^2$ & 0 & 3/8\\
\\
$\langle \bar u u \bar{d} d \rangle $ & $ 1 \langle \bar{u} u \rangle \langle \bar{d} d \rangle$ & 0 & 0\\
$\langle \bar u \gamma^\mu u \bar{d} \gamma_\mu d \rangle $ & $ 0$ & 0 & 5/2\\
$\langle \bar u \sigma^{\mu\nu} u \bar{d} \sigma_{\mu\nu} d \rangle $ & $ 0$ & 0 & 0\\
$\langle \bar u \gamma_5 \gamma^\mu u \bar{d} \gamma_5 \gamma_\mu d \rangle $ & $0$ & 0 & 3/2\\
$\langle \bar u \gamma_5 u \bar{d} \gamma_5 d \rangle $ & $0$ & 0 & 0\\
\\
$\langle \bar u \lambda^A u \bar{d} \lambda^A d \rangle $ & $0$ & 0 & 0\\
$\langle \bar u \gamma^\mu \lambda^A u \bar{d} \gamma_\mu \lambda^A d \rangle $ & $ 0$ & 4/9 & -15/8\\
$\langle \bar u \sigma^{\mu\nu} \lambda^A u \bar{d} \sigma_{\mu\nu} \lambda^A d \rangle $ & $ 0$ & 0 & 0\\
$\langle \bar u \gamma_5 \gamma^\mu \lambda^A u \bar{d} \gamma_5 \gamma_\mu \lambda^A d \rangle $ & $0$ & -2$^*$ & -9/8\\
$\langle \bar u \gamma_5 \lambda^A u \bar{d} \gamma_5 \lambda^A d \rangle $ & $ 0$ & 0 & 0\\
\\
$\langle \bar d d \bar{d} d \rangle $ & $11/12 \langle \bar{d} d \rangle ^2$ & 0 & 0\\
$\langle \bar d \gamma^\mu d \bar{d} \gamma_\mu d \rangle $ & $-1/3 \langle \bar{d} d \rangle ^2$ & 0 & 0\\
$\langle \bar d \sigma^{\mu\nu} d \bar{d} \sigma_{\mu\nu} d \rangle $ & $ -1  \langle \bar{d} d \rangle ^2$ & 0 & 0\\
$\langle \bar d \gamma_5 \gamma^\mu d \bar{d} \gamma_5 \gamma_\mu d \rangle $ & $1/3 \langle \bar{d} d \rangle ^2$ & 0 & 0\\
$\langle \bar d \gamma_5 d \bar{d} \gamma_5 d \rangle $ & $ -1/12 \langle \bar{d} d \rangle ^2$ & 0 & 0\\
\\
$\langle \bar d \lambda^A d \bar{d} \lambda^A d \rangle $ & $ -4/9  \langle \bar{d} d \rangle ^2$ & 0 & 0\\
$\langle \bar d \gamma^\mu \lambda^A d \bar{d} \gamma_\mu \lambda^A d \rangle $ & $-16/9  \langle \bar{d} d \rangle ^2$ & 2/9 & 0\\
$\langle \bar d \sigma^{\mu\nu} \lambda^A d \bar{d} \sigma_{\mu\nu} \lambda^A d \rangle $ & $-16/3  \langle \bar{d} d \rangle ^2$ & 0 & 0\\
$\langle \bar d \gamma_5 \gamma^\mu \lambda^A d \bar{d} \gamma_5 \gamma_\mu \lambda^A d \rangle $ & $ 16/9  \langle \bar{d} d \rangle ^2$ & 1 & 0\\
$\langle \bar d \gamma_5 \lambda^A d \bar{d} \gamma_5 \lambda^A d \rangle $ & $-4/9 \langle \bar{d} d \rangle ^2$ & 0 & 0\\\\
\end{tabular}
\caption{List of 4-quark condensates, their factorized form and their contribution 
to $\omega$ and proton sum rules.}
\end{table}

\section{Electromagnetic Probes in Heavy-Ion Collisions at CERN-SPS Energies}

\begin{figure}[htb]
\vspace*{0.3cm}
\includegraphics[width=5.3cm,angle=-90]{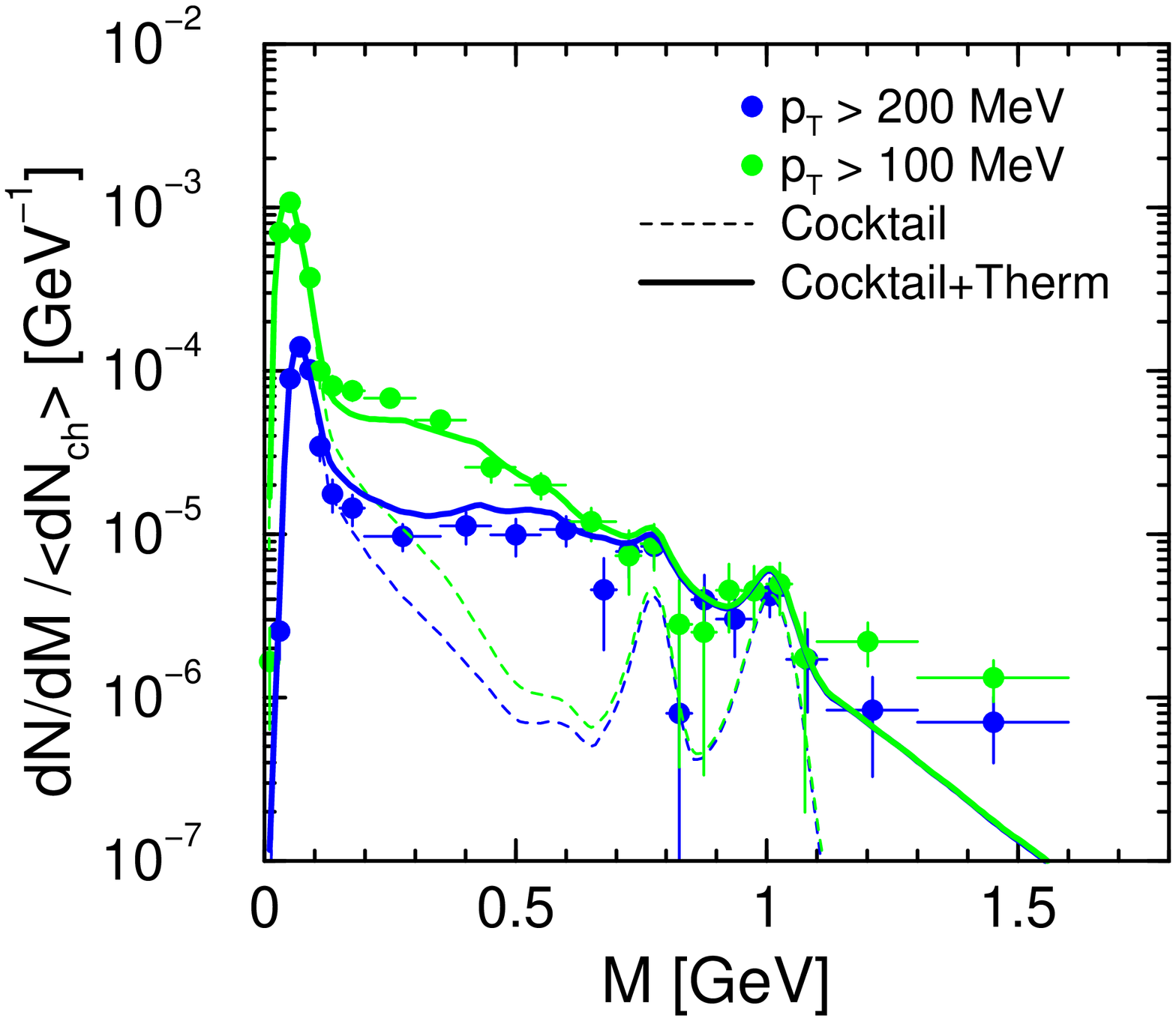}
\hfill
\includegraphics[width=5.3cm,angle=-90]{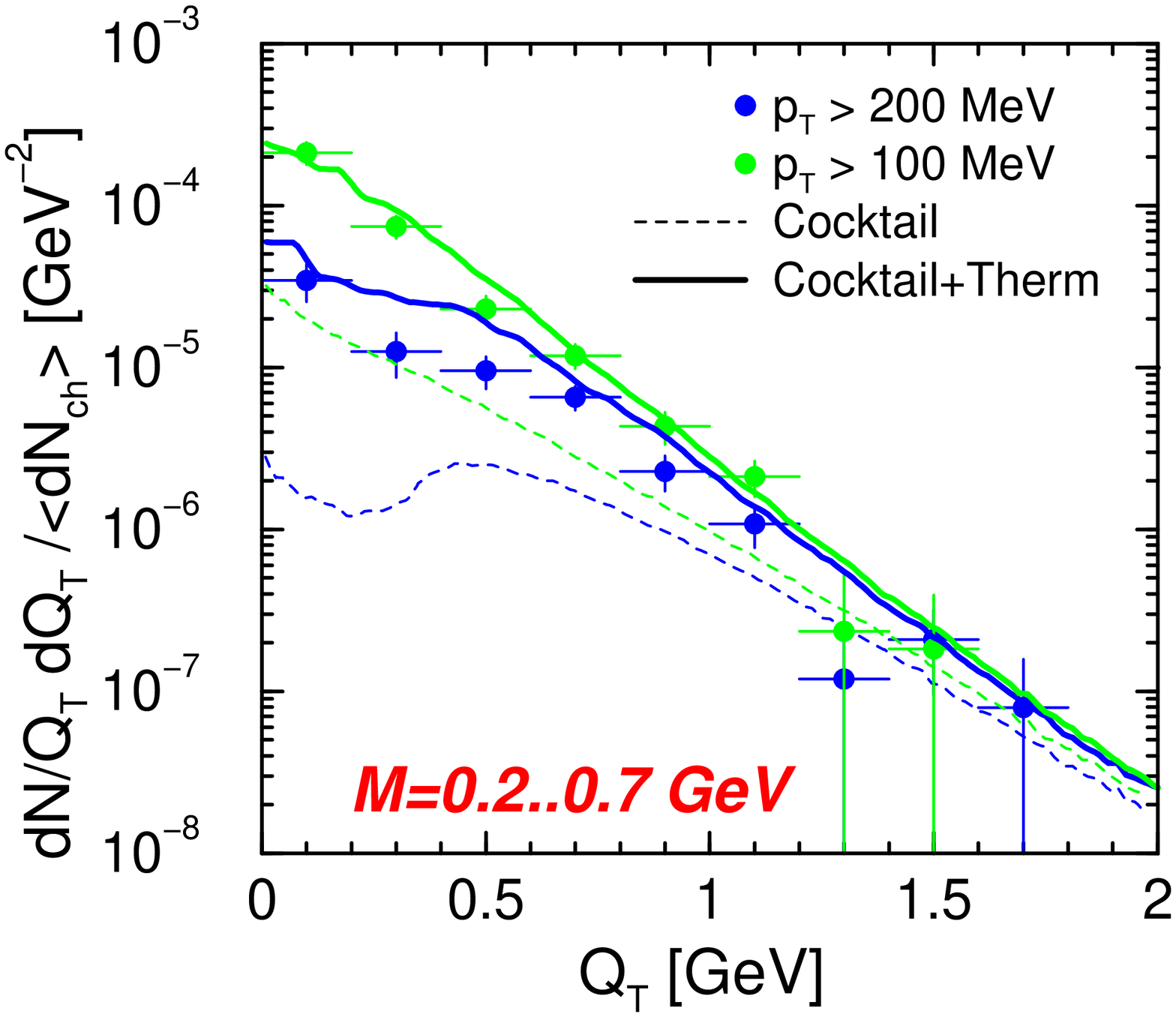}
\vspace*{-0.3cm}
\caption[]{Comparison of our parametrization with new CERES data \cite{CERES}.
Left panel: invariant mass ($M$) spectra, right panel: transverse momentum ($Q_T$) 
spectra for the mass bin $0.2 \cdots 0.7$ GeV.
The minimum single-electron transverse momenta are 200 and 100 MeV/c.
Cocktails are without $\rho$ contribution.} 
\label{fig4}
\end{figure}

To lowest order in temperature, the vector and axialvector correlators
become degenerate at chiral restoration temperature and the vector-axialvector mixing
modifies the dilepton emissivity to a smooth thermal continuum above a certain
duality threshold \cite{Rapp_Wambach}. Extending the duality to soft modes one arrives
at the hypothesis that the emissivity of strongly interacting matter can be parameterized
by the Born term $q \bar q$ annihilation rate, irrespectively whether the matter
is in a confined or deconfined state. 
(The reasoning is that the $e^+ e^-$ rate, in lowest orders in $T$ and $n$,
is related to the inverse reaction
$e^+ e^- \to {\rm hadrons}$ via detailed balance,
i.e.\ ${\rm Im}\Pi_\mu^\mu (s)=-s \alpha \, R(s)$
with $R(s) = \sigma (e^+ e^- \to {\rm hadrons}) / \sigma (e^+ e^- \to \mu^+ \mu^-)$.
The cross features
of $\sigma ( e^+ e^- \to {\rm hadrons})$ in turn can be reproduced by the
Born term for $\sigma ( e^+ e^- \to q \bar q )$ above the $\rho$ resonance region.
The crucial point is therefore the assumption of a medium-caused reshuffling
of electromagnetic strength from the $\rho$ region to smaller energies.)
This ansatz works astonishingly well
\cite{Gallmeister}. Here, we give an update of our parametrization since new
CERES data \cite{CERES} are at our disposal. 
In Fig.~\ref{fig4}, both the invariant mass di-electron spectra
and the transverse momentum spectra are exhibited. In contrast to the parametrization
of previously known data by a single space-time averaged effective radiation
temperature $\langle T \rangle = 170$ MeV, 
the new data within larger phase space coverage enforce the inclusion of
a second thermal radiation component $\langle T \rangle = 120$ MeV 
which may arise from the enlarged sensitivity
to the late stages prior to freeze-out. 

The physics motivation behind such a parametrization is the following observation
\cite{Gallmeister}: Integrating the radiation off an expanding fireball with
volume $V(t)$ and temperature $T(t)$ for certain evolutionary scenarios results
in spectra which can be described by a temperature $\langle T \rangle$ and
an effective space-time volume. Enlarging the phase space coverage by lowering the
minimum single-electron transverse momentum threshold reveals that the final
spectrum appears as superposition of two effective temperatures.   
The new high-statistics NA60 data will shed further light on this issue. 

\begin{figure}[htb]
\vspace*{0.1cm}
\centerline{\includegraphics[width=6.0cm,angle=-90]{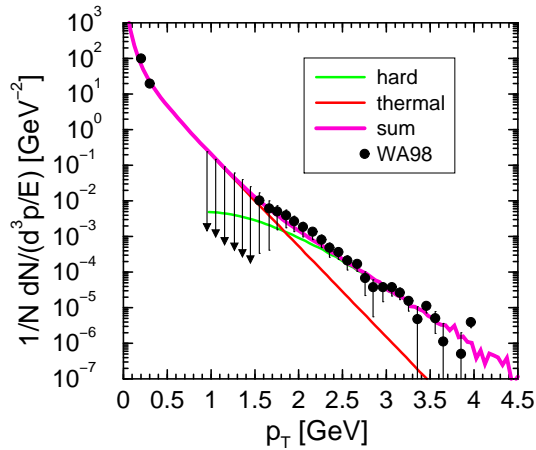}}
\vspace*{0.1cm}
\caption[]{Adopting the same source parametrization as in Fig.~\ref{fig4}
for describing the WA98 photon data \cite{WA98}. The thermal component employs
the complete leading order rate with collinear enhancement factor \cite{AMY}. 
For the calculation of the hard Drell-Yan like component 
see \cite{Gallmeister}.}  
\label{fig5}
\end{figure}

As consistency check one may try to describe the WA98 photon data \cite{WA98} by 
exactly the same source parametrization. Fig.~\ref{fig5} shows that this
new parametrization delivers indeed a better description of the data than the 
previous one \cite{Gallmeister}. 

One might envisage an intimate contact of photon and di-electron rates, as
suggested by (1, 2). The thermal di-electron rate depends on
${\rm Im} \Pi^\mu_\mu (\sqrt{M^2 + \vec Q^2}, \vec Q)$
and is sensitive to the CERES data for $M > 150$ MeV due to strong
contributions of $\pi^0 \to \gamma e^+ e^-$ and $\eta \to \gamma e^+ e^-$
and detector acceptance as well. The photon rate, in contrast, is related to
${\rm Im} \Pi^\mu_\mu (\sqrt{\vec k^2}, \vec k)$, i.e. tests a different
kinematic region. Therefore, one can not establish a direct link of
di-electron and photon rates, in general. In fact, our di-electron rate employs
${\rm Im}\Pi^\mu_\mu = - \frac53 \alpha M^2$, while for the photon rate
we use
${\rm Im}\Pi^\mu_\mu = - 2 \pi \alpha \alpha_s T^2
\left[ \frac12 \log(3/ 4 \pi \alpha_s) + C_{\rm tot} (k_0/T) \right]$
with $C_{\rm tot}$ from equations 1.8 - 1.10 in \cite{AMY}.

\section{Summary}\label{concl}

In summary we argue that the CB-TAPS experiment on $\omega$ photo-production
at nuclei give first hints on a strong density dependence of certain 4-quark
condensates. Such 4-quark condensates contain chirality flipping terms and
may be considered as further order parameters of chiral symmetry restoration.
In contrast to QCD sum rules for the nucleon, the genuine chiral condensate is
unimportant for light vector mesons. Further, basing on vector-axialvector
mixing near the chiral limit we present an upgraded parametrization 
of interesting electromagnetic
probes in heavy-ion collisions at CERN-SPS energies by a  realization and extension
of chiral symmetry restoration as degeneracy of vector and axialvector correlators.

It should be emphasized
that real and virtual photon rates are directly related to the imaginary part
of the current-current correlator, thus linking data to condensates via
QCD sum rules. In such a way the change of the QCD vacuum structure is accessible
at finite temperature and density.
 
\section*{Acknowledgment(s)}
The work is supported by BMBF, GSI, DAAD, and EU.

\vfill\eject
\end{document}